\documentclass[10pt,conference]{IEEEtran}
\usepackage{amsmath,amssymb,amsthm}
\usepackage{stmaryrd}
\usepackage[dvips]{graphicx}
\usepackage{psfrag}
\allowdisplaybreaks[3]

\newcommand{\abs}[1]{\left\lvert#1\right\rvert}

\newtheorem{theorem}{Theorem}
\newtheorem{lemma}{Lemma}
\newtheorem{cor}{Corollary}
\newtheorem{definition}{Definition}

\begin{document}

\title{Performance and Construction of Polar Codes on Symmetric Binary-Input Memoryless Channels}

\author{
\IEEEauthorblockN{Ryuhei Mori}
\IEEEauthorblockA{Department of Systems Science\\
Kyoto University \\
Kyoto, 606-8501, Japan\\
Email: rmori@sys.i.kyoto-u.ac.jp}
\and
\IEEEauthorblockN{Toshiyuki Tanaka}
\IEEEauthorblockA{Department of Systems Science\\
Kyoto University\\
Kyoto, 606-8501, Japan\\
Email: tt@i.kyoto-u.ac.jp}
}

\newcommand{\ANi}[2]{\mathcal{A}_{#1,#2}}
\newcommand{\BNi}[2]{\mathcal{B}_{#1,#2}}
\newcommand{\CNi}[2]{\mathcal{C}_{#1,#2}}
\newcommand{\Ai}[1]{\mathcal{A}_{#1}}
\newcommand{\Bi}[1]{\mathcal{B}_{#1}}
\newcommand{\indices}{\mathcal{I}}
\newcommand{\Bhatt}{Z_N^{(i)}}
\newcommand{\dens}[2]{\mathsf{a}_{#1}^{#2}}

% make the title area
\maketitle

\begin{abstract}
Channel polarization is a method of constructing capacity achieving codes
for symmetric binary-input discrete memoryless channels (B-DMCs)~\cite{arikan2008cpm}.
In the original paper, the construction complexity is exponential in the blocklength.
In this paper, a new construction method for arbitrary symmetric binary memoryless channel (B-MC)
with linear complexity in the blocklength is proposed.
%Furthermore, new upper and lower bounds for the block error probability of polar codes are derived.
%Furthermore, new upper bound for the BEC and lower bound for symmetric B-MC of the block error probability of polar codes are derived.
Furthermore, new upper bound and lower bound of the block error probability of polar codes are derived
for the BEC and arbitrary symmetric B-MC, respectively.
\end{abstract}

\section{Introduction}
Channel polarization, introduced by Ar{\i}kan~\cite{arikan2008cpm}, is a method of constructing capacity achieving codes
for symmetric binary-input discrete memoryless channels (B-DMCs).
Polar codes which are realized by channel polarization require only low encoding and decoding complexity
for achieving capacity.  
Furthermore, it was shown by Ar{\i}kan and Telatar~\cite{arikan2008rcp}
that the block error probability of polar codes is $O(2^{-N^\beta})$ for any fixed $\beta<\frac12$, where $N$ is the blocklength.
It is significantly fast since the block error probability of low-density parity-check (LDPC) codes is polynomial in $N$~\cite{RiU05/LTHC}.
However, in~\cite{arikan2008cpm}, code construction with polynomial complexity is introduced only for the binary erasure channel (BEC)\null.
The main result of this paper is to show code construction with $O(N)$ complexity for arbitrary symmetric binary-input memoryless channel (B-MC)\null.
%Furthermore, new upper and lower bounds for the block error probability of polar codes are derived.
Furthermore, a new upper bound and a lower bound of the block error probability of polar codes are derived
for the BEC and arbitrary symmetric B-MC, respectively.
In Section~\ref{sec:pre}, channel polarization and polar codes introduced in~\cite{arikan2008cpm} are described.
In Section~\ref{sec:cnst}, the construction method for arbitrary symmetric B-MC is shown.
%In Section~\ref{sec:bnd}, new upper and lower bounds of the block error probability of polar code are derived.
In Section~\ref{sec:bnd}, a lower bound of the block error probability of polar codes is derived for arbitrary symmetric B-MC.
In Section~\ref{sec:newup}, a new upper bound of the block error probability of polar codes over the BEC is derived.
%In Section~\ref{sec:bec}, polar codes on the BEC are considered.
In Section~\ref{sec:cal}, some techniques for tightening bounds are discussed.
In Section~\ref{sec:nc}, numerical calculation results are compared with numerical simulation results.
Finally, this paper is concluded in Section~\ref{cncl}.

%Throughout the paper, given a channel $W$, 
%$I(W)$ denotes the mutual information between the input and the output of the channel $W$.

\section{Preliminaries}\label{sec:pre}
\subsection{Channel polarization}
Let the blocklength $N$ be an integer power of 2.  
%Let $u_1^N=(u_1,\,u_2,\,\ldots,\,u_N)$ denote 
%an $N$-dimensional row vector, 
%and let $u_i^j=(u_i,\,u_{i+1},\,\ldots,\,u_j)$ be a subvector of $u_1^N$.  
In~\cite{arikan2008cpm}, Ar\i kan discussed 
channel polarization on the basis of an $N\times N$ matrix $G_N$, 
which he called the generator matrix, defined recursively as 
\begin{equation*}
G_{2^n} := 
R_{2^n}
\left(
F
\otimes
G_{2^{n-1}}
\right)
,\quad
G_2:=F:=
\begin{bmatrix}
1&0\\
1&1
\end{bmatrix}
\end{equation*}
where $\otimes$ denotes Kronecker product
and where $R_{2^n}$ denotes the so-called reverse shuffle matrix, 
which is a permutation matrix.  

For a given B-MC $W:\{0,1\}\to\mathcal{Y}$,
a log-likelihood ratio (LLR) $\log(W(y\mid 0)/W(y\mid 1))$ of $W$ is a sufficient statistic
for estimating input $x\in\{0,1\}$ given output $y\in\mathcal{Y}$.
%Hence, we can assume, without loss of generality, that a log-likelihood ratio is the output of $W$~\cite{RiU05/LTHC}.
Hence, we can associate to $W$ a B-MC $W':\{0,1\}\to\mathbb{R}$ 
with the LLR of $W$ as its output, and $W'$ has the same performance as $W$ under maximum a posteriori (MAP) decoding.
In this paper, we deal with symmetric B-MCs defined as follows.
\begin{definition}
A B-MC $W:\{0,1\}\to\mathcal{Y}$ is said to be symmetric if its associated B-MC $W':\{0,1\}\to\mathbb{R}$
introduced above satisfies
%A real-valued B-MC $W:\{0,1\}\to\mathbb{R}$ is said to be symmetric if it satisfies
$W'(y\mid 0) = W'(-y\mid 1)$.
%\begin{equation*}
%W(x\mid 0) = W(-x\mid 1)\text{.}
%\end{equation*}
\end{definition}
\noindent
Let $I(W)$ denote the capacity between the input and the output of a symmetric B-MC $W$.

We consider communication over a symmetric B-MC $W:\{0,1\}\to\mathbb{R}$.
%We consider communication over a symmetric B-MC $W(y\mid x)$
%where $x$ and $y$ are elements of a set of input alphabets $\{0,1\}$ and a set of output alphabets $\mathbb{R}$, respectively.
%In this paper, we deal with symmetric channels.
Let $u_1^N=(u_1,\,u_2,\,\ldots,\,u_N)$ denote 
an $N$-dimensional row vector, 
and let $u_i^j=(u_i,\,u_{i+1},\,\ldots,\,u_j)$ be a subvector of $u_1^N$.  
Let us consider a vector channel $W_N(y_1^N\mid u_1^N) := W^N(y_1^N\mid u_1^N G_N)$, 
with input $u_1^N \in \{0,1\}^N$ and output $y_1^N\in\mathbb{R}^N$,  
which is obtained by combining $N$ parallel B-DMCs 
$W^N(y_1^N\mid x_1^N):=\prod_{i=1}^NW(y_i\mid x_i)$ via 
the operation $x_1^N=u_1^N G_N$, which should be performed in 
the modulo-2 arithmetic.  
We define subchannels $W_N^{(i)}$ as
\begin{equation*}
W_N^{(i)}(y_1^N, u_1^{i-1}\mid u_i) := \frac1{2^{N-1}}\sum_{u_{i+1}^N}  W_N(y_1^N\mid u_1^N)\text{.}
\end{equation*}
%Let $U_1^N$ be random variable which takes a value $\{0,1\}^N$ with uniform probability
%and $Y_1^N$ be random variable which takes $y_1^N\in\mathbb{R}^N$ with probability $W_N(y_1^N\mid U_1^N)$.
Let $U_1^N\in\{0,1\}^N$ and $Y_1^N\in\mathbb{R}^N$ be random variables 
which follow the joint probability $W_N(y_1^N\mid u_1^N)/2^{N}$.
The mutual information $I(U_1^N ; Y_1^N)$ is split
by applying the chain rule, as 
\begin{align}
I(U_1^N; Y_1^N)&=\sum_{i=1}^N I(U_i\,;\,Y_1^N\mid U_1^{i-1})\nonumber\\
&=\sum_{i=1}^N I(U_i\,;\,Y_1^N,\,U_1^{i-1}) - I(U_i\,;\,U_1^{i-1})\nonumber\\
&=\sum_{i=1}^N I(U_i\,;\,Y_1^N,\,U_1^{i-1})
=\sum_{i=1}^N I(W_N^{(i)})
\text{.}\label{chaini}
\end{align}
%The mutual information in the right-hand side of (\ref{chaini}) can be regarded as mutual information of subchannels.
%The mutual information in the last row of (\ref{chaini}) is mutual information between input and output of subchannels $W_N^{(i)}$.
%Each summand in the last row of (\ref{chaini}) is equal to $I(W_N^{(i)})$.
%Surprisingly, all mutual informations in right-hand side of (\ref{chaini}) take values near zero or one.
Ar\i kan proved the channel polarization property, which states that 
every term in the last line of (\ref{chaini}) 
takes a value near zero or one, 
and that since $I(U_1^N; Y_1^N) = NI(W)$, the approximate numbers of those terms which take values near one and zero 
are $NI(W)$ and $N(1-I(W))$, respectively.
This property suggests the following approach to 
designing a capacity-achieving error-correcting code: 
%designing an efficient error-correcting code: 
%Pick up rows of $G_N$ which correspond to those subchannels
%$W^{(i)}(Y_1^N, U_1^{i-1}\mid U_i) := \sum_{U_{i+1}^N} 2^{-(N-1)} W(Y_1^N\mid U_1^N)$,
Pick up elements of $u_1^N$ which correspond to those subchannels
with high mutual information $I(W_N^{(i)})$, 
and use them as the information bits.  
Non-information bits in $u_1^N$ are clamped to prespecified values.  
The values of the non-information bits are assumed to be all-zero in this paper, 
since they do not affect performance of resulting codes
if the transmitting channel is symmetric~\cite{arikan2008cpm}.
\if0
Evaluation of the values of mutual information $I(U_i\,;\,U_1^{i-1}Y_1^N)$, 
however, requires high computational complexity, so that 
one should consider suboptimal approaches to code construction, i.e., 
to determine what rows of $G_N$ are to be picked up 
to construct an efficient code.  
\fi
%Instead of choosing subchannels with high mutual information $I(U_i\,;\,Y_1^N\mid U_1^{i-1})$,
Instead of choosing subchannels with high mutual information $I(W_N^{(i)})$,
Ar{\i}kan considered another strategy of construction: choosing subchannels with low Bhattacharyya parameters,
which is mentioned later in this section.
%The split of mutual informations of subchannels is origin of a word channel polarization.

\subsection{Decoding}
Ar\i kan considered successive cancellation (SC) decoding
in order to achieve capacity with low complexity.
%in order to define a suboptimal code design procedure.  
%<<<<<<< paper.tex
%In SC decoding, information bits are decoded sequentially from smaller indices.
%If $i$-th bit is not information bit, the decoding result of the $i$-th bit is $0$.
%Otherwise, information bit $u_i$ is decoded by maximum likelihood (ML) like decoding
%which maximizes $W_N^{(i)}(y_1^N,\hat{u}_1^{i-1}\mid u_i)$ 
%instead of true likelihood
%so that for $j>i$, $u_j$ is assumed to be random even if
%the $j$-th bit is not an information bit.
%=======
In SC decoding, decoding results for the non-information bits are set to $0$. 
The information bits are decoded sequentially in the ascending order of their indices, 
via maximum likelihood (ML) decoding of the channel $W_N^{(i)}$. 
%which maximizes $W_N^{(i)}(y_1^N,\hat{u}_1^{i-1}\mid u_i)$ 
%with respect to $u_i\in\{0,\,1\}$.  
%instead of true likelihood
%so that for $j>i$, $u_j$ is assumed to be random even if
%the $j$-th bit is not an information bit.
%>>>>>>> 1.53
%using the whole channel outputs $y_1^N$ 
%and decoding results of information bits $\hat{u}_1^{i-1}$.  
%Otherwise, information bit $u_i$ is decoded as,
More precisely, the decoding result of $i$-th bit is
\begin{equation}
\hat{U}_i(y_1^N,\hat{u}_1^{i-1}) = \mathop{\rm argmax}_{u_i=0,1} W_N^{(i)}(y_1^N, \hat{u}_1^{i-1}\mid u_i)\text{.} \label{dec}
\end{equation}
If the two likelihood values are equal, the decoder determines $0$ or $1$ with probability 1/2.

%where subchannels $W^{(i)}$ are defined as
%\begin{equation*}
%W^{(i)}(y_1^N, u_1^{i-1}\mid u_i) := \frac1{2^{N-1}}\sum_{u_{i+1}^N}  W(y_1^N\mid u_1^N)\text{.}
%\end{equation*}

%A practical method for calculating the likelihood is described in the next section.
%where $i$ is an information bit.
\if0
We consider recursive calculation of log-likelihood ratio (LLR)\null.
From the tree structure of $G_N$, the likelihood in (\ref{dec}) can be calculated by belief propagation (BP) decoding as
\begin{multline*}
L_N^{2i-1}(y_1^N, \hat{u}_1^{2i-2}) =\\
2\tanh^{-1}(\tanh(L_{N/2}^i(y_1^{N/2}, \hat{u}_{1,e}^{2i-2} \oplus \hat{u}_{1.o}^{2i-2})/2)\\
\times\tanh(L_{N/2}^i(y_{N/2+1}^N,\hat{u}_{1,e}^{2i-2})/2))\text{,}
\end{multline*}
\begin{multline*}
L_N^{2i}(y_1^N, \hat{u}_1^{2i-1}) =
L_{N/2}^i(y_{N/2+1}^{N},\hat{u}_{1,e}^{2i-2}) \\
+ (-1)^{\mathbb{I}\{\hat{u}_{2i-1}=1\}}
L_{N/2}^i(y_1^{N/2},\hat{u}_{1,e}^{2i-2} \oplus \hat{u}_{1,o}^{2i-2})\text{.}
\end{multline*}
Although the naive implementation requires $O(N^2)$ complexity, by dynamic programming, the complexity is reduced to $O(N\log N)$.
\fi

\subsection{Upper bound of performance and construction}
%When a subset $\indices\subseteq \{1,2,\dotsc,N\}$ of rows of $G_N$ 
%to be picked up is fixed, 
When a set $\indices\subseteq \{1,2,\dotsc,N\}$ of indices of the information bits is fixed, 
the block error event, denoted by $\mathcal{E}$, of the resulting code with SC decoding
is a union over $\indices$ of the events $\BNi{i}{N}$ that the first bit error occurs 
at the $i$-th bit.
One has 
\begin{align*}
\BNi{i}{N} &= \{\,u_1^N,y_1^N,c_1^N\mid \hat{u}_1^{i-1} = u_1^{i-1}, \hat{U}_i(y_1^N, \hat{u}_1^{i-1}) \ne u_i\,\}\\
 &= \{\,u_1^N,y_1^N,c_1^N\mid \hat{u}_1^{i-1} = u_1^{i-1}, \hat{U}_i(y_1^N, u_1^{i-1}) \ne u_i\,\}\\
&\subseteq \{\,u_1^N,y_1^N,c_1^N\mid \hat{U}_i(y_1^N, u_1^{i-1}) \ne u_i\,\} =: \ANi{i}{N} \label{er}
%&P(B_i)\\
%&\quad= P_{U_1^N,Y_1^N}(\{u_1^N,y_1^N\mid \hat{u}_1^{i-1} = u_1^{i-1}, \hat{U}_i(y_1^N, \hat{u}_1^{i-1}) \ne u_i\})\\
%&\quad= P_{U_1^N,Y_1^N}(\{u_1^N,y_1^N\mid \hat{u}_1^{i-1} = u_1^{i-1}, \hat{U}_i(y_1^N, u_1^{i-1}) \ne u_i\})\\
%&\quad\le P_{U_1^N,Y_1^N}(\{u_1^N,y_1^N\mid \hat{U}_i(y_1^N, u_1^{i-1}) \ne u_i\}) =: P_i \text{.}\label{er}
\end{align*}
where $c_1^N\in\{0,1\}^N$ denote $N$ independent fair coin flips, 
with $c_i$ being used as the decoding result of $u_i$ if 
the two likelihood values for $u_i$ are equal.
In~\cite{arikan2008cpm}, 
$P(\ANi{i}{N})$ is upper bounded by the Bhattacharyya parameter,
\begin{equation*}
\Bhatt := \sum_{y_1^N,u_1^{i-1}}
\sqrt{W_N^{(i)}\left(y_1^N,u_1^{i-1}\mid 0\right)
W_N^{(i)}\left(y_1^N,u_1^{i-1}\mid 1\right)}\text{.}
%Z_i := \sum_{y_1^N,u_1^{i-1}} \biggl(W^{(i)}\left(y_1^N,u_1^{i-1}\mid 0\right)\\
%\times W^{(i)}\left(y_1^N,u_1^{i-1}\mid 1\right)\biggr)^\frac12\text{.}
%\sum_{y_1^N,u_1^{i-1}} \sqrt{P_{Y_1^N,U_1^{i-1}\mid X}(y_1^N,u_1^{i-1}\mid 0) P_{Y_1^N,U_1^{i-1}\mid X}(y_1^N,u_1^{i-1}\mid 1)}\text{.}
\end{equation*}
Hence, the block error probability is upper bounded as
\begin{equation}
P(\mathcal{E}) = \sum_{i\in\indices} P(\BNi{i}{N}) \le \sum_{i\in\indices} P(\ANi{i}{N})\le
\frac12 \sum_{i\in\indices}Z_N^{(i)}\text{.}\label{upb}
\end{equation}
The equality is due to disjointness of $\{\BNi{i}{N}\}$.
The first inequality follows from the above-mentioned inclusion relation between $\ANi{i}{N}$ and $\BNi{i}{N}$.
The last inequality is valid for arbitrary symmetric channels~\cite{RiU05/LTHC}.
In particular, $\Bhatt=2P(\ANi{i}{N})$ if and only if the channel is the BEC\null.
Ar\i kan proposed a method of designing a code in which 
one chooses $\indices$ that minimizes the rightmost side of (\ref{upb}), 
and called the resulting code a polar code.  
In this paper, we propose an alternative code construction strategy 
in which $P(\ANi{i}{N})$ is directly evaluated, instead of $\Bhatt$, 
and $\indices$ that minimizes $\sum_{i\in\indices}P(\ANi{i}{N})$ is chosen.  
We call the codes resulting from our strategy polar codes as well.  

In the rest of this paper,
%unless particular blocklength is considered,
we use the notations $\Ai{i}$ and $\Bi{i}$ instead of $\ANi{i}{N}$ and $\BNi{i}{N}$, respectively,
by dropping the blocklength $N$, 
%unless explicitly specified otherwise.
%except when we specify a different blocklength.
when it is evident from the context.
%when $N$ is an unique blocklength in context.

%In the rest of this paper, we fix a blocklength $N$ and 
%use the notations $\Ai{i}$ and $\Bi{i}$ instead of $\ANi{i}{N}$ and $\BNi{i}{N}$, respectively,
%unless a blocklength is .

\begin{figure}[t]
\psfrag{0}{\tt 000}
\psfrag{1}{\tt 001}
\psfrag{2}{\tt 010}
\psfrag{3}{\tt 011}
\psfrag{4}{\tt 100}
\psfrag{5}{\tt 101}
\psfrag{6}{\tt 110}
\psfrag{7}{\tt 111}
\includegraphics[width=\hsize]{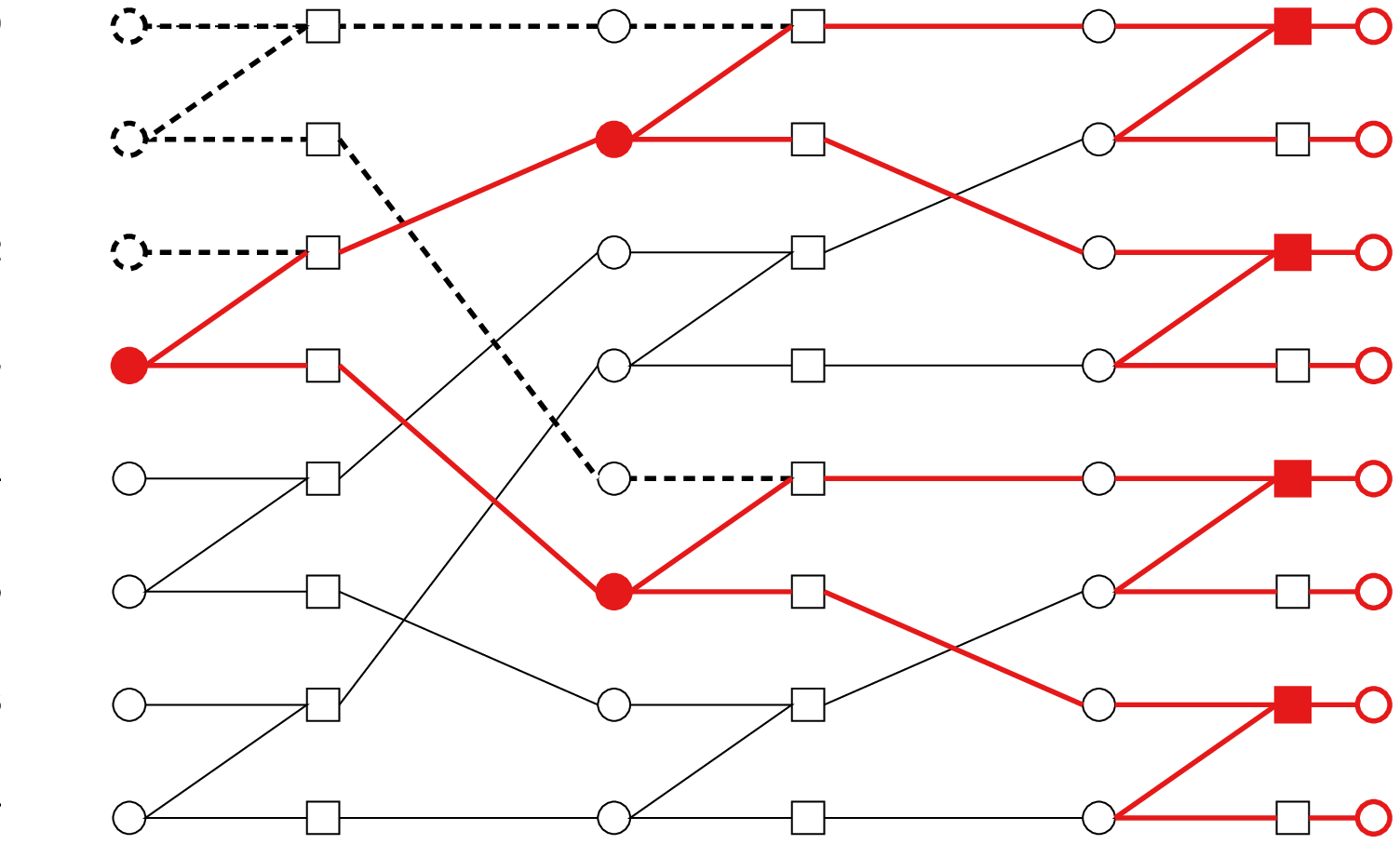}
\caption{The decoding tree for $n=3$, $i=4$. A binary expansion of $(i-1)$ is 011.
Bits $0$ and $1$ in the expansion correspond to check nodes and variable nodes, which are described as filled squares and filled circles, respectively.
Dashed nodes and edges have already been determined to $0$ or $1$ and thus eliminated.
Thin nodes and edges are not useful for decoding for the fourth bit
since thin degree-3 check nodes are connected to a unknown variable node.
The leaf nodes are given messages from a channel.
%The root node is to be decoded by BP decoding on the tree.
}
\label{fig:dec}
\end{figure}

\if0
\begin{figure}[t]
\includegraphics[width=\hsize]{graph.eps}
\caption{Decoding tree of $i=11$. Binary expansion of $11$ is {\tt 1011}.
Binaries {\tt 0} and {\tt 1} correspond to a check and variable nodes, respectively.
The leaf nodes correspond to codeword.}
\label{fig:graph}
\end{figure}
\fi

\section{Construction of Polar codes}\label{sec:cnst}
%\subsection{Belief propagation and density evolution}
%$P_i$ is calculated via density evolution~\cite{RiU05/LTHC}.
%The decoding procedure of polar codes~\cite{arikan2008cpm} is considered as sequential belief propagation (BP) decoding.
%In~\cite{arikan2008cpm}, Ar{\i}kan showed that computation of likelihood in each step of SC decoding (\ref{dec}) could be efficient. 
%In~\cite{arikan2008cpm}, Ar{\i}kan showed that
%$\{P(\Ai{i})\}$ are regarded as the decoding error probability of belief propagation (BP) decoding.
We show in this section that $\{P(\Ai{i})\}$ are regarded as
decoding error probabilities of belief propagation (BP) decoding on tree graphs,
so that they can be evaluated via density evolution.
The Tanner graph of a polar code for $n=3$ is shown in Fig.~\ref{fig:dec}.
Let us consider $i$-th step of SC decoding.
Since $u_1^{i-1}$ have been either determined as non-information bits or decoded in previous steps, 
the edges incident to these variable nodes are eliminated.
Since $u_{i+1}^N$ do not affect the characteristics of the channel $W_N^{(i)}$, 
the degree-3 check nodes connected to them do not work in this stage. 
Hence, these check nodes and the edges incident to them are eliminated.
Similarly, degree-3 check nodes incident to undetermined degree-1 variable nodes are also eliminated recursively.
The resulting decoding graph for $u_i$ is tree-like, as shown in Fig.~\ref{fig:dec}.
%When $u_1^{i-1}$ is determined, the remain tanner graph become shown like Fig.~\ref{fig:dec} due to following reasons.
%When $u_1^{i-1}$ is determined, the remain tanner graph become shown like Fig.~\ref{fig:dec} due to following reasons.
%Since $u_1^{i-1}$ is known, edges incident to them are also known.
%We assume known edges to be vanished. 
%Further, variable nodes connect to known edges and
%edges incident to known variable node or degree $1$ check nodes are also known.
%%Known edges increase recursively.
%Similarly, we consider unknown edges.
%Since $u_{i+1}^N$ is unknown, degree $3$ check nodes incident to them do not work.
%the likelihood is calculated by belief propagation (BP)\null.
%Ar{\i}kan future consider dynamic programming in order to reduce whole computational complexity of SC decoding
%from $O(N^2)$ to $O(N\log N)$.
%Since in this paper, we consider performance of SC decoding but do not consider decoding algorithm,
%Hence, the ML decision (\ref{dec}) is regarded as belief propagation (BP) decoding on the tree graph.
%Hence, the ML decision (\ref{dec}) can be implemented by belief propagation (BP) decoding on the tree graph.
Hence, the ML decision (\ref{dec}) can be implemented by BP decoding on the tree graph.
The probability $P(\Ai{i})$ is therefore regarded as
the error probability of the root node of the tree graph via BP decoding,
where leaf nodes have messages of the channel.
%The tree graphs consist of variable nodes and check nodes. 
Assume that the binary expansion of $(i-1)$ is $b_n\dotsc b_1$,
then nodes at depth $t$ of the tree graph are check nodes
and variable nodes if $b_t=0$ and $b_t=1$, respectively, as shown in Fig.~\ref{fig:dec}\footnote{%
In counting the depth we omit nodes in the tree with degree 2, 
because messages of BP are passed through such nodes unprocessed.}. 

An LLR for $i$-th bit, defined as
%\begin{equation*}
%L_N^{(i)}(y_1^N,\hat{u}_i^{i-1}) := \log\frac{W_N^{(i)}(y_1^N,\hat{u}_1^{i-1}\mid 0)}{W_N^{(i)}(y_1^N,\hat{u}_1^{i-1}\mid 1)}
%\end{equation*}
$L_N^{(i)}(y_1^N,\hat{u}_i^{i-1}) := \log(W_N^{(i)}(y_1^N,\hat{u}_1^{i-1}\mid 0)/W_N^{(i)}(y_1^N,\hat{u}_1^{i-1}\mid 1))$
is calculated recursively as
\begin{multline*}
L_N^{(2i-1)}(y_1^N, \hat{u}_1^{2i-2}) \\
=2\tanh^{-1}(\tanh(L_{N/2}^{(i)}(y_1^{N/2}, \hat{u}_{1,e}^{2i-2} \oplus \hat{u}_{1.o}^{2i-2})/2)\\
\times\tanh(L_{N/2}^{(i)}(y_{N/2+1}^N,\hat{u}_{1,e}^{2i-2})/2))
\end{multline*}
\begin{multline*}
L_N^{(2i)}(y_1^N, \hat{u}_1^{2i-1}) =
L_{N/2}^{(i)}(y_{N/2+1}^{N},\hat{u}_{1,e}^{2i-2}) \\
%+ (-1)^{\mathbb{I}\{\hat{u}_{2i-1}=1\}}
+ (-1)^{\hat{u}_{2i-1}}
L_{N/2}^{(i)}(y_1^{N/2},\hat{u}_{1,e}^{2i-2} \oplus \hat{u}_{1,o}^{2i-2})
\end{multline*}
where $\hat{u}_{1,e}^i$ and $\hat{u}_{1,o}^i$ denote subvectors
which consist of elements of $\hat{u}_1^i$ with even and odd indices, respectively,
and where $\oplus$ denotes modulo-2 addition.
%Although the naive implementation requires $O(N^2)$ complexity, by dynamic programming, the complexity is reduced to $O(N\log N)$.
The above updating rules are originally derived by Ar{\i}kan~\cite{arikan2008cpm}.

It is well known in the field of LDPC codes
that the error probability of the root node of a tree graph after message passing decoding is calculated via density evolution.
%In order to use this method, we restrict the class of channels to symmetric channels 
%whose output range is a subset of $\mathbb{R}$.
\if0
In order to use density evolution, we restrict the class of channels to symmetric channels.
For any channels, LLR of channel output is sufficient statistics.
Hence, we regard the channel output as LLR\null.

%For introducing the 
%whose output range is a subset of $\mathbb{R}$.
\begin{definition}
A real-valued B-MC $W:\{0,1\}\to\mathbb{R}$ is said to be symmetric if it satisfies
$W(x\mid 0) = W(-x\mid 1)$.
%\begin{equation*}
%W(x\mid 0) = W(-x\mid 1)\text{.}
%\end{equation*}
\end{definition}
%The real-valuedness assumption does not exclude discrete channels since arbitrary finite or countable sets can be regarded as subsets of $\mathbb{R}$.
%The real-valuedness assumption is not restriction for any channels since LLR of channel output is sufficient statistics
%and can be regarded as the channel output without loss of generality~\cite{RiU05/LTHC}.
\fi
%For the analysis of the error probability of symmetric channels, without loss of generality,
%it is assumed that the all-zero message is transmitted
%and that all non-information bits $U_i$, $i\in\indices^c$, are fixed to be zero.
For the analysis of the error probability of symmetric channels, without loss of generality,
it is assumed that the all-zero message is transmitted.
The following theorem for symmetric B-MC is a consequence of a well-known result in~\cite{RiU05/LTHC}
and also obtained for the BEC by Ar{\i}kan~\cite{arikan2008cpm}.

\begin{theorem}\label{cstrct}
For a symmetric B-MC which has a density $\mathsf{a}_W$ of LLR,
it holds that
%\begin{equation*}
$
P(\Ai{i}) = \mathfrak{E}(\dens{N}{i})
$
%\end{equation*}
where 
\begin{equation*}
%\mathfrak{E}_1(a) := \lim_{\epsilon\to +0}\left( \int_{-\infty}^{-\epsilon} a(x) \mathrm{d}x + \frac1{2}\int_{-\epsilon}^{+\epsilon} a(x) \mathrm{d}x\right)\text{,}
%\mathfrak{E}(a) := \int_{-\infty}^{0} 2^{-\mathbb{I}\{x=0\}}a(x) \mathrm{d}x \text{,}
\mathfrak{E}(\mathsf{a}) := \lim_{\epsilon\to +0} \left(\int_{-\infty}^{-\epsilon} \mathsf{a}(x) \mathrm{d}x
 + \frac12\int_{-\epsilon}^{+\epsilon} \mathsf{a}(x)\mathrm{d}x\right)\text{,}
\end{equation*}
\begin{align*}
\dens{2N}{2i} &= \dens{N}{i} \star \dens{N}{i}\text{,}&
\dens{2N}{2i-1} &= \dens{N}{i} \boxast \dens{N}{i}\text{,}&
\dens{1}{1} &= \mathsf{a}_{W}
\end{align*}
and where $\star$ and $\boxast$ denote the convolutions of LLR density functions,
which are defined in~\cite{RiU05/LTHC},
corresponding to variable and check nodes, respectively.
%\begin{align*}
%a_{2N}^{2i} &= a_N^i \star a_N^i\text{,}\\
%a_{2N}^{2i-1} &= a_N^i \boxast a_N^i\text{,}\\
%a_1^1 &= a_{W}\text{.}
%\end{align*}
\end{theorem}
\noindent
On the basis of the availability of $P(\Ai{i})$s assured by Theorem~\ref{cstrct}, 
we propose the following code construction procedure: 
%The following code construction procedure can be considered.
Choose $\indices$ which minimizes
\begin{equation}\label{upai}
\sum_{i\in\indices}P(\Ai{i})\text{,}
\end{equation}
subject to $\abs{\indices}=NR$.
The block error probability of the resulting codes also decays like $O(2^{-N^\beta})$ for any $\beta<\frac12$
as in~\cite{arikan2008rcp}, 
since the upper bound of the block error probability, 
given in terms of Bhattacharyya parameters in~\cite{arikan2008rcp}, 
is also an upper bound of the block error probability of the codes 
constructed via the proposed method.  

%\subsection{Complexity of construction}
In~\cite{arikan2008cpm}, complexity of code construction on the BEC is explained as $O(N\log N)$.
However, Theorem~\ref{cstrct} states that the complexity of code construction, 
not only for the BEC but also for an arbitrary symmetric B-MC, is $O(N)$.  
To see this, 
let $\chi(N)$ denote the complexity of calculation of $\{\dens{N}{i}\}_{i=1,\dotsc,N}$
where the complexities of computations of $\star$ and $\boxast$ are considered to be constant.
Then, it is evaluated as
\begin{equation*}
\chi(N) = N + \chi\left(\frac{N}{2}\right) = N + \frac{N}{2} + \frac{N}{4} + \dotsb + 1 = O(N)\text{.}
\end{equation*}
Since the complexity of selecting the $NR$-th smallest $P(\Ai{i})$ is $O(N)$ even in the worst case~\cite{blum:tbs},
the complexity of code construction is $O(N)$.

We would like to note that larger $N$ requires higher-precision representation 
of messages for reliable SC decoding and density evolution computations.  
In this regard,
the complexity of SC decoding discussed in~\cite{arikan2008cpm}
and
the complexity of construction mentioned above
should be
understood as referring to
the number of the arithmetics of LLRs in BP and
the number of the convolution operations in density evolution,
respectively, not mentioning their precision.
%In this regard, the complexity results mentioned above should be 
%In this regard, the complexity of construction mentioned above should be 
%understood as referring to the numbers of the convolution operations in density evolution
%as well as the complexity of SC decoding discussed in~\cite{arikan2008cpm} should be
%understood as referring to the numbers of the arithmetics of real numbers,
%not mentioning their precision.  
%and of arithmetic operations involved in SC decoding, not mentioning their precision.  
%In fact, however, the required number of quantization level in density evolution has to grow as $N$ increases.
%Hence, constant cost assumption of the convolutions is not valid.
%If the number of quantization level in density evolution is fixed, the construction is not exact.
%The inexactness of the construction due to constant level quantization in density evolution corresponds to
%inexactness of SC decoding due to constant level quantization of real numbers.
In practice, use of finite-sized binning in density evolution may lead to imprecise upper bounds of 
the block error probabilities, which, however, still provide upper bounds 
relevant to SC decoding with the same quantization as the binning scheme.  
%Due to the above discussion, at the construction,
%Strictly speaking,
%the number of the convolutions of probability density functions is $O(N)$
%as well as the number of arithmic of real numbers is $O(N\log N)$ at SC decoding.

%??
%Strictly speaking, at the construction, the number of the convolutions of probability density functions is $O(N)$
%as well as the number of arithmic of real numbers is $O(N\log N)$ at SC decoding.

%Following Ar{\i}kan, we consider that the complexity of multiplication of real numbers is constant
%as well as the complexity of the convolutions is constant.

%This blocklength-dependness of the cost of the convolutions corresponds to that
%of the cost of multiplication of real numbers.

%On the other hand, 
%in order to determine the order of indices for SC decoding,
%one has to sort indices with $O(N\log N)$ complexity.

\section{Lower bound of the block error probability for arbitrary symmetric B-MC}\label{sec:bnd}
%\section{Lower bound of the block error probability of particular polar codes}\label{sec:bnd}
%\section{New bounds of the block error probability of particular polar codes}\label{sec:bnd}
%\subsection{Inclusion-exclusion principle}
%On the one hand, 
%the upper bound (\ref{upai}) of the block error probability of polar codes on the BEC yields poor results.
%In particular, it exceeds one near the capacity as observed in~\cite{arikan2008cpm} for the BEC.
%On the other, no lower bound of the block error probability has been known. 
%In this section, we introduce lower bound for a given choice of information bits $\indices$.
%%In this section, we introduce new upper and lower bounds for a given choice of information bits $\indices$.
%We use the following fundamental lemma.

To the authors' knowledge, no lower bound of the block error probability of polar codes has been known. 
In this section, we introduce a lower bound for a given choice of information bits $\indices$.
We use the following fundamental lemma.

\begin{lemma}\label{bi}
$\bigcup_{i\in\indices} \Bi{i} = \bigcup_{i\in\indices} \Ai{i}$ .
%\begin{equation*}
%\bigcup_{i\in\indices} \Bi{i} = \bigcup_{i\in\indices} \Ai{i}\,\text{.}
%\end{equation*}
\end{lemma}
\begin{IEEEproof}
The direction $\subseteq$ is trivial.
Assume an event $v$ belongs to $\Ai{i}$.
If $\hat{u}_1^{i-1} = u_1^{i-1}$, $v$ belongs to $\Bi{i}$.
Otherwise, $v$ belongs to $\Bi{j}$ for some $j<i$ which belongs to $\indices$.
\end{IEEEproof}
Recalling ${\cal E}=\bigcup_{i\in\indices}\Bi{i}$, 
it immediately follows that $P({\cal E})= P(\bigcup_{i\in\indices}\Ai{i})$ holds.
%From this lemma, the block error probability of polar codes is expressed as the probability of union of $\Ai{i}$s over all information bits,
%i.e.,
%$P(\bigcup_{i\in\indices}\Bi{i})= P(\bigcup_{i\in\indices}\Ai{i})$.
The events $\{\Ai{i}\}$ are easier to deal with than $\{\Bi{i}\}$.
Several bounds which use probabilities concerning $\{\Ai{i}\}$ are considered in what follows.

First, via Boole's inequality, the following lower bound is obtained for any $\mathcal{S}\subseteq \indices$
\begin{multline}
P\left(\bigcup_{i\in\indices}\Ai{i}\right)
%=\max_{\mathcal{S}\subseteq \indices} P\left(\bigcup_{i\in \mathcal{S}}\Ai{i}\right)\\
\ge P\left(\bigcup_{i\in \mathcal{S}}\Ai{i}\right)\\
%\ge \max_{\mathcal{S}\subseteq \indices}
\ge
\sum_{i\in\mathcal{S}}P\left(\Ai{i}\right)
-\sum_{(i,j)\in\mathcal{S}^2,i< j}P\left(\Ai{i}\cap \Ai{j}\right)\text{.}\label{lowbnd}
\end{multline}
%Maximization of the right hand side of the above inequality is difficult.
%In order to evaluate the lower bound (\ref{lowbnd}), calculation of $P(\Ai{i}\cap \Ai{j})$s are required.

Maximization of the lower bound (\ref{lowbnd}) with respect to $\mathcal{S}$ is difficult
since it is equivalent to the Max-Cut problem, which is NP-hard~\cite{caprara2008cqp}.
However, without strict optimization, one can obtain practically accurate lower bounds for some rates and channels.

In order to obtain the lower bound~\eqref{lowbnd}, evaluations of probabilities of intersections of two $\Ai{i}$s are required.
%To obtain new upper and lower bounds, evaluation of $P(A_{i_1}\cap\dotsm\cap A_{i_t})$ is required.
%In the following subsection, the method called \textit{joint density evolution} is introduced for the evaluations.
%The method called \textit{joint density evolution} is introduced for this purpose.
For this purpose, we introduce a method which we call the \textit{joint density evolution}.
Let $(X_1, Y_1)$ and $(X_2, Y_2)$ denote pairs of random variables which independently follow
$\mathsf{a}(x,y)$ and $\mathsf{b}(x,y)$, respectively.
The convolution $\mathsf{a}\mathop{\star\star} \mathsf{b}$ is defined as the joint density function of messages $(X, Y)$
where $X=X_1+X_2$ and $Y=Y_1+Y_2$.
Similarly, the convolutions $\mathsf{a}\mathop{\star\boxast}\mathsf{b}$ is defined as the joint density function of messages $(X,Y)$
where $X=X_1+X_2$ and $Y=2\tanh^{-1}(\tanh(Y_1/2)\tanh(Y_2/2))$.
% and where $(X_1,Y_1)$ and $(X_2,Y_2)$ are pairs of random variables
%which independently follow $a(x,y)$ and $b(x,y)$, respectively.
The other convolutions $\mathsf{a}\mathop{\boxast\star}\mathsf{b}$ and $\mathsf{a}\mathop{\boxast\boxast}\mathsf{b}$
are also defined in the same way.
%Details of the definitions are omitted for lack of space.

\begin{theorem}\label{a2}
For a symmetric B-MC which has a density $\mathsf{a}_W$ of LLR,
the joint density $\dens{N}{i,j}$ of LLR for $i$-th bit and $j$-th bit after BP decoding is calculated recursively as
\if0
\begin{align*}
P(\Ai{i}\cap \Ai{j})&=\mathfrak{E}_2(a_N^{i,j})\text{,}\\
P(\Ai{i}^c\cap \Ai{j})&=\mathfrak{E}_3(a_N^{i,j})\text{,}\\
P(\Ai{i}\cap \Ai{j}^c)&=\mathfrak{E}_4(a_N^{i,j})\text{,}\\
P(\Ai{i}^c\cap \Ai{j}^c)&=\mathfrak{E}_5(a_N^{i,j})\text{,}
\end{align*}
where
\begin{align*}
\mathfrak{E}_2(a) &:= \int_{-\infty}^{0}\int_{-\infty}^{0} 2^{-\mathbb{I}\{x=0\}-\mathbb{I}\{y=0\}} a(x,y) \mathrm{d}x\mathrm{d}y\text{,}\\
\mathfrak{E}_2(a) &:= \int_{-\infty}^{0}\int_{-\infty}^{0} 2^{-\mathbb{I}\{x=0\}-\mathbb{I}\{y=0\}} a(x,y) \mathrm{d}x\mathrm{d}y\text{,}\\
\mathfrak{E}_2(a) &:= \int_{-\infty}^{0}\int_{-\infty}^{0} 2^{-\mathbb{I}\{x=0\}-\mathbb{I}\{y=0\}} a(x,y) \mathrm{d}x\mathrm{d}y\text{,}\\
\mathfrak{E}_2(a) &:= \int_{-\infty}^{0}\int_{-\infty}^{0} 2^{-\mathbb{I}\{x=0\}-\mathbb{I}\{y=0\}} a(x,y) \mathrm{d}x\mathrm{d}y\text{,}
\end{align*}
\fi
\begin{align*}
\dens{2N}{2i,2j}&=\dens{N}{i,j} \;\star\star\;  \;\dens{N}{i,j}\text{,}&
\dens{2N}{2i,2j-1}&=\dens{N}{i,j} \,\star\boxast\,  \;\dens{N}{i,j}\text{,}\\
\dens{2N}{2i-1,2j}&=\dens{N}{i,j} \,\boxast\star\,  \;\dens{N}{i,j}\text{,}&
\dens{2N}{2i-1,2j-1}&=\dens{N}{i,j} \boxast\boxast  \;\dens{N}{i,j}\text{,}\\
\dens{1}{1}(x,y)&=\delta(x-y)\mathsf{a}_W(x)
%a_1^{1,1}(x,y)&=\begin{cases}
%a_W(x),&\text{if } x=y\text{,}\\
%0,&\text{otherwise.}
%\end{cases}
\end{align*}
where $\delta(x)$ denotes the Dirac delta function.
\end{theorem}
\noindent
The probabilities $P(\Ai{i}\cap \Ai{j})$, $P(\Ai{i}^c\cap \Ai{j})$, $P(\Ai{i}\cap \Ai{j}^c)$ and $P(\Ai{i}^c\cap \Ai{j}^c)$ are
calculated by appropriate integrations of the joint density $\dens{N}{i,j}$.
%Although further extensions of joint density evolution to higher order are also possible straightforwardly,
%they are omitted due to lack of space.
Extensions of joint density evolution to higher order joint distributions are also possible straightforwardly.

For the BEC, density evolution has only to evolve expectations of erasure probabilities~\cite{RiU05/LTHC}.
%In this section, joint density evolutions for the BEC are shown.
Correspondingly, joint density evolution for the BEC is much simpler than that for a general symmetric B-MC, as follows.

\begin{cor}\label{cor:bec}
For the BEC with erasure probability $\epsilon$,
\begin{multline*}
\dens{N}{i,j}(x,y) = p_{N}^{i,j}(0,0)\delta(x)\delta(y) + p_{N}^{i,j}(0,1)\delta(x)\delta_\infty(y)\\
+ p_{N}^{i,j}(1,0)\delta_\infty(x)\delta(y) + p_{N}^{i,j}(1,1)\delta_\infty(x)\delta_\infty(y)
\end{multline*}
%\begin{align*}
%P(\Ai{i}\cap \Ai{j}) &= \frac14 p_N^{i,j}(0,0)\text{,}\\
%P(\Ai{i}^c\cap \Ai{j}) &= \frac12 p_N^{i,j}(1,0) + \frac14 p_N^{i,j}(0,0)\text{,}\\
%P(\Ai{i}\cap \Ai{j}^c) &= \frac12 p_N^{i,j}(0,1) + \frac14 p_N^{i,j}(0,0)\text{,}\\
%P(\Ai{i}^c\cap \Ai{j}^c) &= p_N^{i,j}(1,1) + \frac12 \left(p_N^{i,j}(1,0) + p_N^{i,j}(0,1)\right)\\
%&\quad + \frac14 p_N^{i,j}(0,0)
%P(\Ai{i}'\cap \Ai{j}') &= p_N^{i,j}(0,0)\text{,}\\
%P(\Ai{i}'^c\cap \Ai{j}') &= p_N^{i,j}(1,0)\text{,}\\
%P(\Ai{i}'\cap \Ai{j}'^c) &= p_N^{i,j}(0,1)\text{,}\\
%P(\Ai{i}'^c\cap \Ai{j}'^c) &= p_N^{i,j}(1,1) 
%\end{align*}
where
\begin{align*}
p_{2N}^{2i,2j}(0,0) &= p_{N}^{i,j}(0,0)^2\text{,}\\
p_{2N}^{2i,2j}(0,1) &= p_{N}^{i,j}(0,1)^2 + 2p_{N}^{i,j}(0,0) p_{N}^{i,j}(0,1)\text{,}\\
p_{2N}^{2i,2j}(1,0) &= p_{N}^{i,j}(1,0)^2 + 2p_{N}^{i,j}(0,0) p_{N}^{i,j}(1,0)\text{,}\\
p_{2N}^{2i,2j}(1,1) &= 1-p_{2N}^{2i,2j}(0,0)\\
&\quad - p_{2N}^{2i,2j}(0,1) - p_{2N}^{2i,2j}(1,0)\text{,}
\end{align*}
\begin{align*}
p_{2N}^{2i,2j-1}(0,0) &= p_{N}^{i,j}(0,0)^2 + 2p_{N}^{i,j}(0,0)p_{N}^{i,j}(0,1)\text{,}\\
p_{2N}^{2i,2j-1}(0,1) &= p_{N}^{i,j}(0,1)^2\text{,}\\
p_{2N}^{2i,2j-1}(1,1) &= p_{N}^{i,j}(1,1)^2 + 2p_{N}^{i,j}(1,1) p_{N}^{i,j}(0,1)\text{,}\\
p_{2N}^{2i,2j-1}(1,0) &= 1-p_{2N}^{2i,2j-1}(0,0)\\
&\quad - p_{2N}^{2i,2j-1}(0,1) - p_{2N}^{2i,2j-1}(1,1)\text{,}
\end{align*}
\begin{align*}
p_{2N}^{2i-1,2j}(0,0) &= p_{N}^{i,j}(0,0)^2 + 2p_{N}^{i,j}(0,0)p_{N}^{i,j}(1,0)\text{,}\\
p_{2N}^{2i-1,2j}(1,0) &= p_{N}^{i,j}(1,0)^2\text{,}\\
p_{2N}^{2i-1,2j}(1,1) &= p_{N}^{i,j}(1,1)^2 + 2p_{N}^{i,j}(1,1) p_{N}^{i,j}(1,0)\text{,}\\
p_{2N}^{2i-1,2j}(0,1) &= 1-p_{2N}^{2i-1,2j}(0,0)\\
&\quad - p_{2N}^{2i-1,2j}(1,0) - p_{2N}^{2i-1,2j}(1,1)\text{,}
\end{align*}
\begin{align*}
p_{2N}^{2i-1,2j-1}(0,1) &= p_{N}^{i,j}(0,1)^2 + 2p_{N}^{i,j}(0,1)p_{N}^{i,j}(1,1)\text{,}\\
p_{2N}^{2i-1,2j-1}(1,0) &= p_{N}^{i,j}(1,0)^2 + 2p_{N}^{i,j}(1,0)p_{N}^{i,j}(1,1)\text{,}\\
p_{2N}^{2i-1,2j-1}(1,1) &= p_{N}^{i,j}(1,1)^2\text{,}\\
p_{2N}^{2i-1,2j-1}(0,0) &= 1-p_{2N}^{2i-1,2j-1}(0,1)\\
&\quad - p_{2N}^{2i-1,2j-1}(1,0) - p_{2N}^{2i-1,2j-1}(1,1)\text{,}
\end{align*}
\begin{align*}
p_1^{1,1}(0,0) &= \epsilon\text{,}&
p_1^{1,1}(0,1) &= 0\text{,}\\
p_1^{1,1}(1,0) &= 0\text{,}&
p_1^{1,1}(1,1) &= 1-\epsilon
\end{align*}
and where $\delta_\infty(x)$ denotes the Dirac delta function of unit mass at infinity.
\end{cor}
%Note that the probabilities regarding $\{\Ai{i}\}$ are obtained from the probabilities concerned with $\{\Ai{i}'\}$,
%e.g. $P(\Ai{i}\cap \Ai{j}) = P(\Ai{i}'\cap\Ai{j}')/4$.
Higher-order joint probabilities such as $P(\Ai{i}\cap \Ai{j}\cap \Ai{k})$ are calculated recursively
by tracking real vectors of an appropriate dimension 
in a way similar to that described in Corollary~\ref{cor:bec}.  
%Since the number of cases is also $8$, the formulas are omitted.

%Complexity of computations of all $a_N^{i,j}$ for $\{i,j\}\subseteq \{1,\dotsc,N\}$, $i\ne j$ is $O(N^2)$ as $N$ tends to infinity.
Complexity of computations (as measured in numbers of convolution operations) of all $\dens{N}{i,j}$s is $O(N^2)$ as $N$ increases.
%Similarly, complexity of computations of all $s$-joint densities $a_N^{i_1,\dotsc,i_s}$ for $\{i_1,\dotsc,i_s\}\in\{1,\dotsc,N\}^s$,
%$\abs{\{i_1,\dotsc,i_s\}}=s$ is $O(N^s)$ as $N$ tends to infinity.
Similarly, complexity of computations of all $s$-joint densities $\dens{N}{i_1,\dotsc,i_s}$s is $O(N^s)$ as $N$ increases.
%which means that the complexity grows exponentially in $s$ since the dimension of densities is $s$.
%On the other hand, the complexity grows exponentially in $s$ since the dimension of the densities is $s$ and
%since the number of combinations $\binom{N}{s}$ is exponential in $s$.
On the other hand, the complexity grows exponentially in $s$ since the dimension of the densities is $s$.

\section{New upper bound of the block error probability for the BEC}\label{sec:newup}
The upper bound (\ref{upai}) of the block error probability of polar codes may yield poor results.
In particular, it exceeds one near the capacity, as observed in~\cite{arikan2008cpm} for the BEC\null.
In this section, a new upper bound which does not exceed one is derived for the BEC\null.
%The upper bound (\ref{upai}) of the block error probability of polar codes yields poor results.
%In particular, it exceeds one near the capacity as observed in~\cite{arikan2008cpm} for the BEC.
%In order to obtain a tighter upper bound, we consider Bhattacharyya parameter bound.
%In other words, a performance of codes over the BEC with erasure probability
%which is equal to the Bhattacharyya parameter of the original channel is considered,
%in which case the obtained performance is always worse than the performance over the original
%channel~\cite{arikan2008cpm,RiU05/LTHC}.
%In the following, only the BEC is considered as the transmitting channel.
%Hence, we assume that the transmitting channel is the BEC\null.
%Hence, we give here an upper bound only for the BEC\null.
For the BEC, covariances among complements of $\{\Ai{i}\}$, denoted by $\{\Ai{i}^c\}$, are always positive.
\begin{lemma}
For the BEC,
$P(\bigcap_{i\in\indices} \Ai{i}^c) \ge \prod_{i\in\indices} P(\Ai{i}^c)$,
for any $\indices\subseteq\{1,\dotsc,N\}$.
%$P(\Ai{i}^c\cap \Ai{j}^c) \ge P(\Ai{i}^c)P(\Ai{j}^c)$.
%\begin{equation*}
%P(\Ai{i}^c\cap \Ai{j}^c) \ge P(\Ai{i}^c)P(\Ai{j}^c)\text{.}
%\end{equation*}
\end{lemma}
\begin{IEEEproof}[Outline of proof]
An event $\Ai{i}^c$ is expressed as $\bigcap_k \left\{e_{i,k}\nsubseteq \mathcal{F}\right\}$,
where $e_{i,k}$ is an error pattern of erasure messages for $i$-th bit, 
and where $\mathcal{F}$ is a set of indices of erasure messages.
%If an indicator function $\mathbb{I}_{\Ai{i}^c}(\cdot)$ of outputs of the channel increases or decreases, $\mathbb{I}_{\Ai{j}^c}(\cdot)$,
%$j\ne i$, does not decrease or increase, respectively.
%The inequality is obtained recursively from Corollary~\ref{cor:bec} which is shown in Section~\ref{sec:bec} below.
\end{IEEEproof}
Using this property, the block error probability is upper bounded simply by
\if0
\begin{multline}
P\left(\bigcup_{i\in\indices}\Ai{i}\right)=
1-P\left(\bigcap_{i\in\indices}\Ai{i}^c\right)\\
= 1-\prod_{k=1}^{\abs{\indices}} P\left(A_{i_k}^c \Bigm | \bigcap_{m=1}^{k-1} A_{i_m}^c\right)\text{,}\label{upbnd0}
\end{multline}
In the above equation, the conditional probability decrease by eliminating any conditions, i.e.
\begin{equation*}
P\left(A_{i_k}^c \Bigm|  \bigcap_{m=1}^{k-1} A_{i_m}^c\right)
\ge P\left(A_{i_k}^c \Bigm|  \bigcap_{m=1}^{k-2} A_{i_m}^c\right)
%P\left(A_{i_k}^c\mid A_{i_1}^c,\dotsc,A_{i_{k-1}}^c\right) \ge P\left(A_{i_k}^c\mid A_{i_1}^c,\dotsc,A_{i_{k-2}}^c\right)
\text{,}
\end{equation*}
since.
Thus, (\ref{upbnd0}) is upper bounded by
\begin{equation}
1-\prod_{k=1}^{\abs{\indices}} P\left(A_{i_k}^c\right)\text{,}\label{upbnd1}
\end{equation}
Furthermore, more accurately (\ref{upbnd0}) is upper bounded by
\begin{equation}
1-\prod_{k=1}^{\abs{\indices}} P\left(A_{i_k}^c\mid A_{p(i_{k})}^c\right)\text{,}\label{upbnd2}
\end{equation}
\fi
%\begin{equation*}
$1-\prod_{i\in\indices} P\left(\Ai{i}^c\right)$.
%\end{equation*}
%where $\{i_1,\dotsc,i_{\abs{\indices}}\}=\indices$.
%Furthermore, more accurately (\ref{upbnd0}) is upper bounded by
Furthermore, it is more accurately upper bounded by
\begin{equation}
1-\prod_{i\in\indices} P\left(\Ai{i}^c\mid \Ai{p(i)}^c\right)\label{upbnd2}
\end{equation}
where $p(i)\in\indices$ corresponds to a parent node of $\Ai{i}$ in a spanning tree of a perfect graph
which has nodes corresponding to indices in $\indices$.
$P(\Ai{i}^c\mid \Ai{p(i)}^c)$ is calculated via joint density evolution.
In order to tighten the upper bound (\ref{upbnd2}), the maximum weight directed spanning tree should be chosen
from the perfect directed graph whose edges have weights $P(\Ai{i}^c\mid \Ai{j}^c)$, where $i$ and $j$ are sink and source nodes of the directed edge, respectively,
%where each edge in the perfect directed graph has a weight as $P(\Ai{i}^c\mid \Ai{j}^c)$ where $i$ and $j$ are sink and source nodes of the edge, respectively
like Chow-Liu tree~\cite{chow1968adp}.

%For general symmetric B-MC $W$ with Bhattacharyya parameter $Z$, the error probability of BP on tree graph
%over $W$ is better than that over BEC$(Z)$~\cite{arikan2008cpm}, \cite{RiU05/LTHC}.
%Hence, the upper bound~\eqref{upbnd2} for BEC$(Z)$ is an upper bound for $W$. 
%Then, one obtains an upper bound which is always smaller than 1.

%\section{Polar codes on the binary erasure channel}\label{sec:bec}
%\section{New upper bound and several techniques for calculations of bounds}\label{sec:cal}
\section{Techniques for tightening bounds}\label{sec:cal}
%The bounds which are derived in the previous sections can further be refined and simplified 
In this section, some techniques for obtaining tighter bounds of $P(\bigcup_{i\in\mathcal{I}} A_i)$ are shown.
%The bounds \eqref{upai} and \eqref{lowbnd} can further be refined and simplified 
%if the channel is the BEC.
%The upper bound (\ref{upai}) of the block error probability of polar codes on the BEC yields poor results.
%
%On the BEC, polar codes have several simple properties.
%The bounds which are derived in the previous sections can further be refined and simplified 
%if the channel is the BEC.
The first one is applicable to polar codes over the BEC. 
In this case, the LLR $L_N^{(i)}(y_1^N,\hat{u}_j^{i-1})$ for $i$-th bit in SC decoding, 
when the all-zero information is transmitted, is either zero or infinity.
Let $\Ai{i}'$ be the event that the LLR for $i$-th bit is zero.  
We consider the events of erasure $\{\Ai{i}'\}$ rather than $\{\Ai{i}\}$ for simplicity.
%We consider partial ordering on $\{0,1\}^N$.
%We first define partial ordering on $\{0,1\}^N$.
We first define partial ordering on $\{1,\dotsc,2^n\}$.
\begin{definition}
For $i,j \in \{1,\dotsc,2^n\}$, $i\prec j$ if and only if $t$-th bit of binary expansion of $j-1$ is one
when $t$-th bit of binary expansion of $i-1$ is one for any $t\in\{1,\dotsc,n\}$. 
%For $i,j \in \{1,\dotsc,2^n\}$, $i\prec j$ if and only if $t$-th element of $j-1$ is one if $t$-th element of $i-1$ is one. 
%We also say $i<j$ if and only if $i\le j$ and $i\ne j$.
\end{definition}
\noindent
The following theorem is useful for reducing time complexity of calculations of the bounds.

\begin{lemma}\label{min}
%if there exists $j\in\indices$ such that $j\prec i$ and $j\ne i$ then $P(\Bi{i})=0$.
$j\prec i\Longleftrightarrow \Ai{i}' \subseteq \Ai{j}'$.
\end{lemma}
\begin{IEEEproof}
If a variable node outputs an erased message, a check node with the same input as the variable node
outputs an erased message. Hence if $v\in \Ai{i}'$ then $v\in \Ai{j}'$.
The proof of the other direction is also obvious and is omitted.
%For another direction, if $j\nprec i$, there exists $v\in \Ai{i}$ such that $v\notin \Ai{j}$.
\end{IEEEproof}

\if0
\begin{cor}
%Assume $\mathcal{B}$ is a subset of a set of information bits $\indices$
%such that $\land_{i\in\mathcal{B}}i \in \indices$ and
%$\lor_{i\in\mathcal{B}}i \in \indices$, then
For $\mathcal{B}\subseteq\{0,\dotsc,N-1\}$,
\begin{align*}
\bigcup_{i\in \mathcal{B}} \Ai{i} &\subseteq A_{\land_{i\in\mathcal{B}}i}\text{ ,}\\
\bigcup_{i\in \mathcal{B}} \Ai{i}^c &\subseteq A_{\lor_{i\in\mathcal{B}}i}^c\text{ .}
%P\left(\bigcup_{i\in \mathcal{B}} \Bi{i}~\bigg|~ B_{\land_{i\in\mathcal{B}}i}^c\right)&=0\\
%P\left(\bigcap_{i\in \mathcal{B}} \Bi{i}^c\right) &\ge P\left(B_{\land_{i\in\mathcal{B}}i}^c\right)\\
%P\left(B_{\lor_{i\in\mathcal{B}}i}~\bigg|~\bigcup_{i\in \mathcal{B}} \Bi{i}^c\right) &= 0
%P\left(\bigcap_{i\in \mathcal{B}} B_i\right) &\ge P\left(B_{\lor_{i\in\mathcal{B}}i}\right)
\end{align*}
\end{cor}
\fi

\begin{theorem}\label{th:min}
The block erasure probability of polar codes of information bits $\indices$ is
$P(\bigcup_{i\in M(\indices)}\Ai{i}')$
where $M(\indices)$ denotes the set of minimal elements of $\indices$ with respect to $\prec$.
\end{theorem}
\begin{IEEEproof}
From Lemma~\ref{min},
$\bigcup_{i\in\indices} \Ai{i}' = \bigcup_{i\in M(\indices)} \Ai{i}'$.
%From Lemma~\ref{bi} and Lemma~\ref{min},
%$\bigcup_{i\in\indices} B_i 
%= \bigcup_{i\in\indices} \Ai{i}
%= \bigcup_{i\in M(\indices)} \Ai{i}
%$.
\end{IEEEproof}
\noindent
From this Theorem, we have only to consider the set of minimal elements $M(\indices)$
for the block erasure probability of polar codes over the BEC, 
which can be used to tighten bounds.  
%The following corollary is a consequence of Theorem~\ref{a2} and Lemma~\ref{min}.

%\section{Performance on the binary erasure channel}

%For general B-MC, similar result using the partial order $\prec$ only for low digits is obtained.
For polar codes over a general symmetric B-MC, the following result similar to Theorem~\ref{th:min} is obtained.
\begin{theorem}\label{thm:gai}
For integers $0 \le k\le n$,
and $0 \le i\le 2^{n-k}-1$
\begin{equation*}
%P(\ANi{2^ki+1}{2^n}\cup\ANi{2^ki+2}{2^n}\cup\dotsm\cup\ANi{2^k(i+1)}{2^n})\\
P(\ANi{2^ki+1}{2^n}\cup\dotsm\cup\ANi{2^k(i+1)}{2^n})
 = 1 - (1-P(\ANi{i+1}{2^{n-k}}))^{2^k}
\end{equation*}
\end{theorem}
\noindent
Proof is omitted for lack of space.
Although the joint probability $P(\ANi{2^ki+1}{2^n}\cup\dotsm\cup\ANi{2^k(i+1)}{2^n})$ can be calculated
via joint density evolution,
Theorem~\ref{thm:gai} allows us to calculate it 
more efficiently via density evolution for depth-$(n-k)$ trees and a few arithmetics.

From Theorem~\ref{thm:gai}, one can efficiently obtain a tighter upper bound than \eqref{upai}
%From Theorem~\ref{thm:gai}, one can obtain tighter bounds using \eqref{upai} and \eqref{lowbnd}
%The block error event $\bigcup_{i\in\mathcal{I}}\Ai{i}$ is considered as an union of sets
%which are expressed as $\ANi{2^ki+1}{2^n}\cup\dotsm\cup\ANi{2^k(i+1)}{2^n}$.
by decomposing the block error event as $\mathcal{E} =\bigcup_{i\in\mathcal{I}}\Ai{i} = \bigcup_{j\in\mathcal{J}}\mathcal{C}_{j}$,
where each $\mathcal{C}_j$ is expressed as $\ANi{2^ki+1}{2^n}\cup\dotsm\cup\ANi{2^k(i+1)}{2^n}$.
%which is union of $\Ai{i}$s among indices of size power of 2
%with appropriate alignment
%as a block
%in the upper bound \eqref{upai} and the lower bound \eqref{lowbnd}.
For example, if we choose $\mathcal{I}=\{4, 6, 7, 8\}$ as information bits for a polar code with $N=8$,
one obtains an upper bound $P(\Ai{4})+P(\Ai{6})+P(\Ai{7}\cup\Ai{8})$, which is tighter than
$P(\Ai{4}) + P(\Ai{6}) + P(\Ai{7}) + P(\Ai{8})$.

\section{Numerical calculations and simulations}\label{sec:nc}
\begin{figure}[t]
\psfrag{x}{$\epsilon$}
\psfrag{y}{$P_B$}
\psfrag{up1}{\small (\ref{upai})}
\psfrag{up2}{\small (\ref{upbnd2})}
\psfrag{oo1}{\small (\ref{lowbnd})}
\psfrag{sim}{simulation}
\includegraphics[width=\hsize]{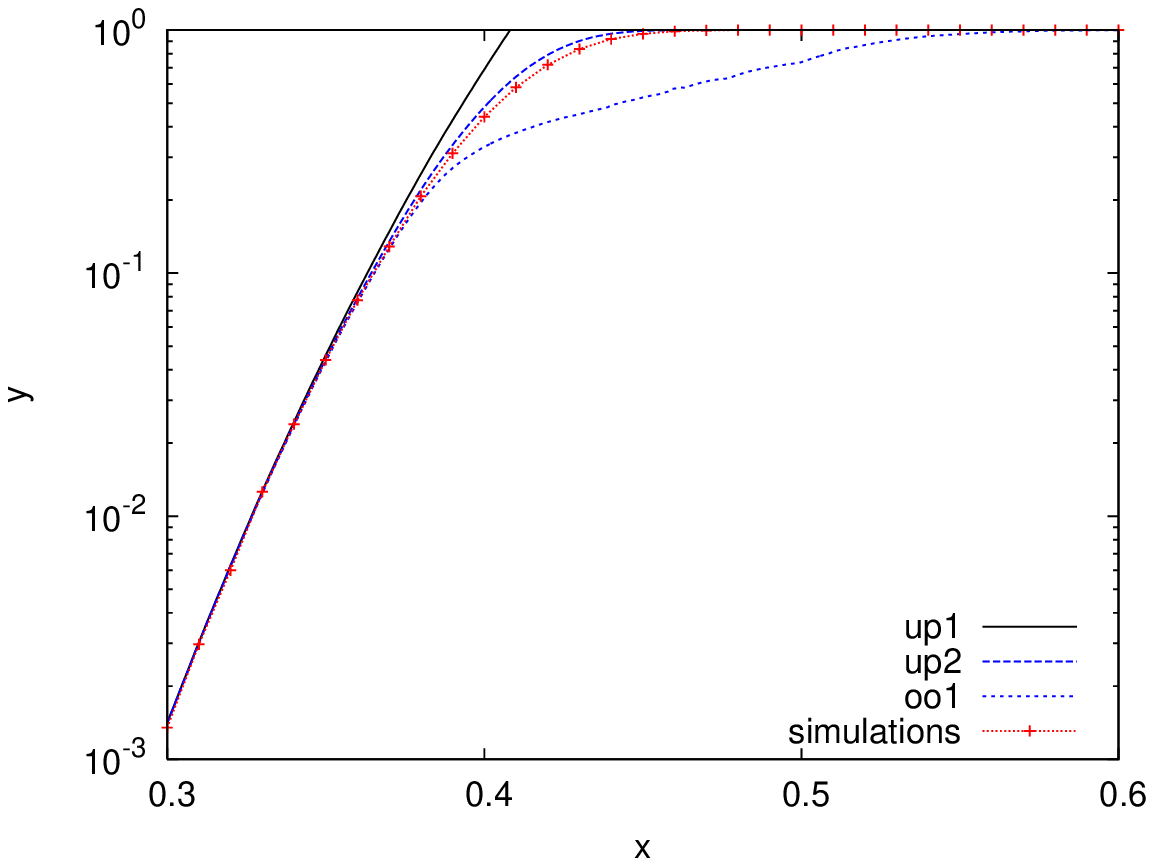}
\caption{Calculation results of upper bounds (\ref{upai}), (\ref{upbnd2}) and lower bound (\ref{lowbnd}).
Rate is $0.5$. Blocklength is $1024$.}
\label{figcal}
\end{figure}

In this section, numerical calculation results are compared with numerical simulation results.
\if0
Numerical simulation results are shown in Fig.~\ref{figsim}.
Codes which are optimized for each $\epsilon$ by the criterion~(\ref{upai}) are simulated over the BEC\null.
Coding rate is $0.5$ and blocklengths are $2^n$ where $n=4,\dotsc,10$.
Below the Shannon threshold $\epsilon=0.5$, the block error probability approaches $0$ and
above $0.5$, the block error probability approaches $1$.
Since the blocklength is not sufficiently large,
slightly below $0.5$, the block error probability increases as the blocklength increases.
\fi
Figure~\ref{figcal} shows calculation results of the upper bounds~(\ref{upai}), (\ref{upbnd2}) and the lower bound~(\ref{lowbnd})
of block erasure probability.
Coding rate is $0.5$ and blocklength is $1024$.
Only the minimal elements of information bits are considered in view of Theorem~\ref{th:min} for calculation of these bounds.
Although we optimized the upper bound (\ref{upbnd2}) and the lower bound (\ref{lowbnd}) only approximately,
the lower bound is very close to the upper bound for $\epsilon$ below $0.4$.
Our new upper bound is always smaller than $1$ and closer to the simulation results, whereas the union bound exceeds $1$ when $\epsilon>0.407$.

\section{Conclusion and future works}\label{cncl}
The construction method of polar codes for symmetric B-MCs with complexity $O(N)$ is shown.
New upper and lower bounds for the block error probability of particular polar codes and
the method of joint density evolution are derived.
The method and the bounds are also applicable to generalized polar codes~\cite{korada2009pcc}.

Computing higher-order joint distributions and deriving other bounds (e.g., Boole's inequality with higher-order terms) are future works.
%Obtaining bounds using higher order joint distributions is a future work.

\if0
\begin{figure}[t]
\psfrag{x}{$\epsilon$}
\psfrag{y}{$P_B$}
\includegraphics[width=\hsize]{polar_sim.eps}
\caption{Simulation results of polar codes over the BEC\null. Rate is $0.5$. Blocklengths are $2^n$ where $n=4,\dotsc,10$.}
\label{figsim}
\end{figure}
\fi

\section*{Acknowledgment}
%TT acknowledges support through Grant-in-Aid for Scientific Research on Priority Areas (No. 18079010) from MEXT, Japan.
TT acknowledges support of the Grant-in-Aid for Scientific Research
on Priority Areas (No.~18079010), MEXT, Japan.

%\enlargethispage{-1em}
\bibliographystyle{IEEEtran}
\bibliography{IEEEabrv,ldpc}

\end{document}